# Atomistic Mechanisms of Temperature-Dependent Ion Track Formation in Gallium Nitride under Swift Heavy Ion Irradiation

*Jiayu Liang[1], Shaowei He[1], Wenlong Liao[2,3], Tan Shi[1], Hang Zang[1], Yonghong Li[1], Xiaojun Fu[2,3], Chuanjian Yao[2,3], Chaohui He[1\*], Jianan Wei[2,3\*] and Huan He[1\*]*

[1] School of Nuclear Science and Technology, Xi'an Jiaotong University, Xi'an, 710049, China

[2] Academy of Chips Technology, China Electronics Technology Group Corporation, Chongqing, 400060, China

[3] National Key Laboratory of Integrated Circuits and Microsystems, Chongqing, 400060, China

[*]E-mail: hechaohui@xjtu.edu.cn, weijianan93@163.com & huanhe@xjtu.edu.cn

Abstract：

The radiation tolerance of gallium nitride under extreme conditions is critical for its deployment in next-generation electronic and optoelectronic devices, yet the microscopic mechanisms governing swift heavy ion induced damage at elevated temperatures remain poorly understood. Therefore, this study employs a coupled approach including two temperature model and molecular dynamics simulations to resolve the entire processes of ion track generation induced by swift heavy ions irradiation across a wide temperature range. A temperature-driven morphological transition of ion tracks, evolving from discontinuous segments to continuous tracks composed of isolated nanobubbles, and ultimately to fully continuous channels is observed. Under lower electronic stopping loss of 430 MeV Kr irradiation, increasing temperature significantly enhances track visibility, enlarges track radii and promotes nanobubble formation. For higher electronic stopping conditions of 1171 MeV Ta irradiation, continuous ion tracks consisting of discontinuous nanobubbles (~1.5 nm radius) emerge already at 300 K, followed by a thermally activated transition into continuous channels with further radial expansion. At the atomic scale, SHI irradiation induces decomposition of wurtzite GaN into Ga clusters and $N_2$ molecules along the ion trajectory, with Ga-rich regions and recrystallized wurtzite phases accumulating near bubble interfaces, while $N_2$ preferentially segregates within bubble cores. Additionally, zincblende nanodomains nucleate around ion tracks and exhibit strong spatial correlation with radiation-induced dislocation networks, particularly screw dislocations, providing potential pathways for leakage current and increased susceptibility to single-event burnout. These findings establish a comprehensive atomistic picture of temperature-dependent track evolution in GaN and offer critical insights into radiation-induced degradation mechanisms in wide-bandgap semiconductor devices.

1 Introduction

As a third-generation semiconductor material, gallium nitride (GaN) exhibits exceptional performance and promising applications in radiation-hardened electronics such as light-emitting diodes (LEDs), laser diodes, and high electron mobility transistors (HEMTs) due to its excellent properties. Compared to silicon-based devices, GaN HEMTs not only achieve higher efficiency, operate at higher voltages, and enable faster switching transitions, but also demonstrate superior thermal stability and greater tolerance to ionizing radiation[1–5]. These characteristics make GaN the preferred choice for device fabrication in extreme environments such as space and high-energy experimental facilities.

Although GaN is generally regarded as highly resistant to radiation, significant radiation-induced performance degradation is still observed in practical applications. Single-event effect (SEE) experiments on GaN-based devices indicate a prominent sensitivity to swift heavy ions (SHIs) [6–8]. Regarding the characteristics of single-event failure, Hu *et al.* observed performance degradation including increased on-resistance and reduced breakdown voltage in GaN HEMTs after SHI irradiation. Through experimental characterization such as transmission electron microscopy (TEM), they identified line-shaped latent tracks extending from the AlGaN/GaN heterojunction channel to the GaN buffer layer[9,10]. Current simulations based on Technology Computer Aided Design (TCAD) tools can replicate radiation-induced damage patterns by modeling charge distribution and transport. However, the conclusions drawn from it are typically macroscopic, attributing device failure to current jumps caused by localized charge accumulation from carrier drift[11,12]. At the microscopic mechanism level, Sequeira *et al.* identified strong recrystallization as a core mechanism underlying the resistance of GaN to intense ionizing radiation and established an empirical relationship between the electronic energy loss ($S_e$) values of SHIs and the molten/amorphous cross-section[13,14]. Mahfuz *et al.* validated that the density and diameter of nanovoids in GaN increase with rising ion fluence[15].

Nonetheless, SHI irradiation generates high-density electron-hole pairs, triggering a sharp localized temperature surge. During this process, the formation of various defects causes material damage, leading to severe performance degradation and potentially catastrophic device failure. Arto *et al.* reported that Joule heating induced by SHI irradiation under sufficient bias voltage in SiC Schottky diodes can generate permanent lattice defects[16]. Xiao *et al.* observed the activation of latent damage in SiC MOSFETs under post-irradiation gate stress, with localized Joule heating resulting in complex defect structures including surface amorphization and extended bulk damage tracks[17]. Through multiscale simulations, Zainutdinov *et al.* highlighted suppressed recrystallization in SiC at elevated temperatures, with damage severity increasing at higher temperatures[18]. These studies collectively emphasize that temperature is not merely an auxiliary factor but rather a dominant driving force in the evolution of radiation-induced defects. For GaN devices, some studies indicate that the single-event burnout (SEB) threshold decreases significantly with increasing operating voltage, as the accompanying temperature surge becomes more intense at higher voltages[19–22]. However, the underlying microscopic physical mechanisms by which high temperatures drive the severe degradation or catastrophic failure of GaN devices under SHI impact remain unclear. This fundamental knowledge gap significantly hinders the widespread deployment of GaN devices in extreme aerospace environments.

2 Methods

To investigate the unresolved role of transient high temperatures induced by SHI irradiation, we have adopted a multi-scale simulation approach that combines two temperature model (TTM) with MD simulations to elucidate the temperature-dependent evolution of latent ion track morphology and microscopic damage mechanisms in GaN. In this work, two typical SHIs including 430 MeV Kr and 1171 MeV Ta ions were employed to simulate the SHI irradiation, corresponding to $S_e$ values of 18.2 and 40.2 keV/nm, respectively. The irradiation was performed centrally along the [0001] crystallographic direction in a simulation cell measuring 30 nm × 30 nm × 50 nm. The initial simulation temperatures were set from 300 to 1800 K. Additional simulation details are provided in Section 2 of the supplementary materials.

3 Result

The cross-sectional schematics of GaN under 430 MeV Kr and 1171 MeV Ta irradiation at 300K at three representative times are firstly presented in Figures 1(a) and 1(b), respectively: the initial stage at 1 ps, the maximum expansion of the amorphous zone at 20 ps, and the stabilized state at 750 ps. These visualizations clearly depict the dynamic evolution of atomic displacement and recrystallization in GaN following the thermal spike induced by SHI irradiation. The thermal spike forms a cylindrical amorphous zone centered around the ion trajectory[13,14]. In the initial stage of long-term MD simulations, the extremely high temperature at the ion track center drives continuous radial expansion of the amorphous zone until it reaches its maximum radius. Under the effect of the thermal bath layer, the GaN system exchanges heat with the external environment, leading to a gradual temperature decrease. Recrystallization initiates from the outer edge of the amorphous zone toward the ion track center. When the system stabilizes at the set temperature, the recrystallization process is essentially complete, resulting in an amorphous zone with a specific radius. At 300 K, the final amorphous core radius induced by 430 MeV Kr irradiation is approximately 0.1 nm, while that induced by 1171 MeV Ta irradiation reaches approximately 1.5 nm. This observation indicates a positive correlation between the $S_e$ values of the incident SHI and the resulting amorphous zone radius.

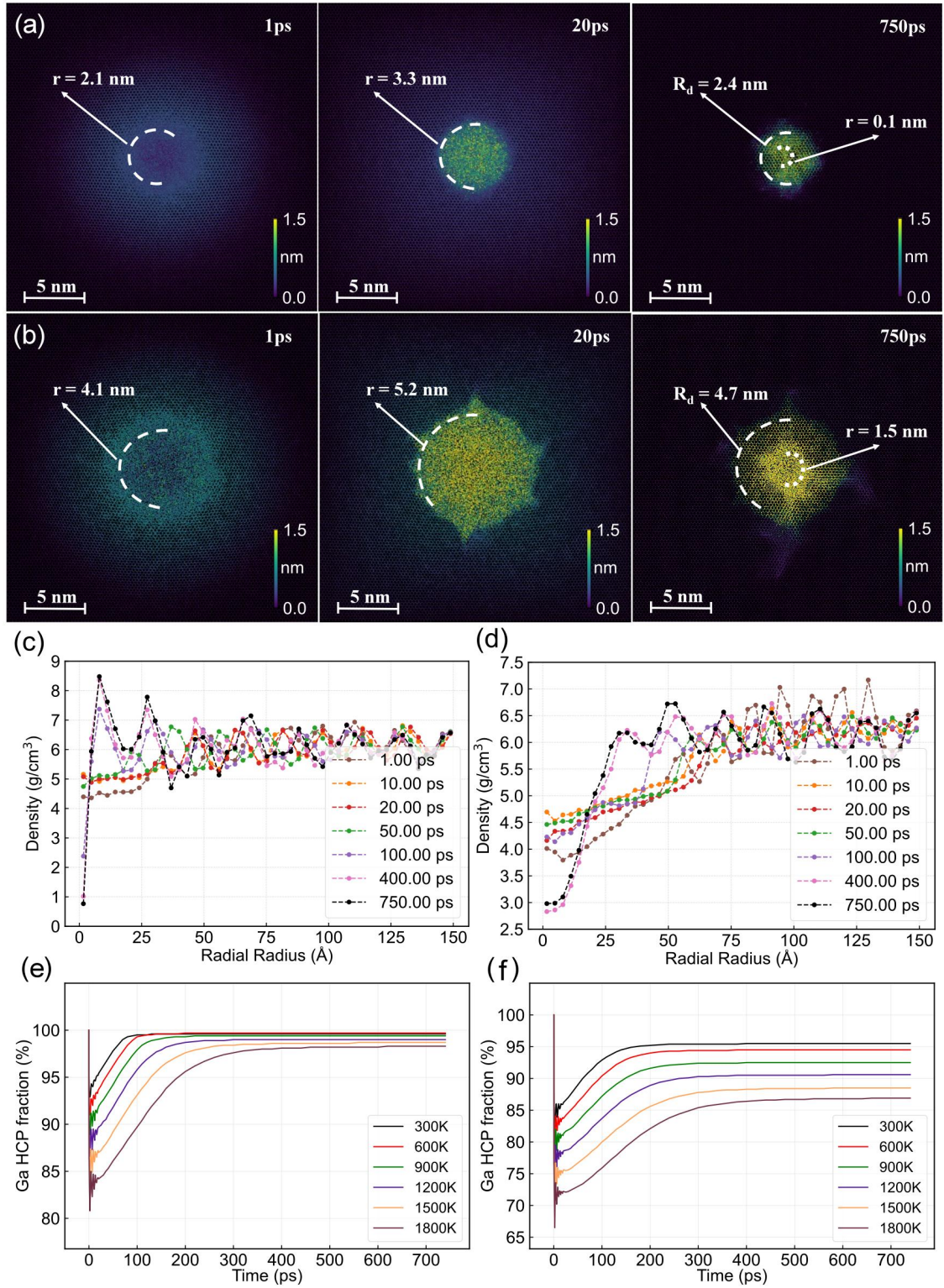

**FIG. 1.** The cross-sectional schematics (25 nm thickness) of GaN at 200 nm depth along the ion incidence path under (a) 430 MeV Kr and (b) 1171 MeV Ta irradiation at 300 K. Here, $r$ denotes the amorphous molten region radius, and $R_d$ indicates the atomic displacement radius. The color scheme represents the displacement magnitude of atoms relative to original positions. Radial density profiles at representative key times under (c) 430 MeV Kr and (d) 1171 MeV Ta irradiation at 300 K. The HCP structure (Ga) fractions in GaN under (e) 430 MeV Kr and (f) 1171 MeV Ta irradiation at different temperatures.

The evolution of radial density distribution confirms this melt-recrystallization process, as displayed in Figures 1(c) and 1(d). Within the first 20 picoseconds (ps), the radial distance corresponding to the solid-state density of GaN (6.15 g/cm$^3$) gradually increases, reflecting radial expansion of the amorphous region at the ion track center. After 20 ps, this radial distance gradually decreases, demonstrating the shrinkage of the amorphous zone and the onset of recrystallization. By the final stage, the recrystallization process is largely complete, with the amorphous region reaching a final configuration. Additionally, the atomic density at the radial center ($r = 0$) exhibits a significant decrease, reflecting that SHI irradiation leaves a highly amorphous core with substantial structural defects. To further elucidate the atomic-scale structural evolution following SHI impact, the fractions of Ga hexagonal close-packed (HCP) structure within the wurtzite GaN under 430 MeV Kr and 1171 MeV Ta irradiation at different temperatures are shown in Figures 1(e) and 1(f), respectively. The high temperature generated instantly upon SHI irradiation melts the wurtzite lattice, causing a sharp drop in the HCP fraction of Ga atoms. Due to the competing effects of molten zone expansion and edge recrystallization, persistent fluctuations appear in the early stage. After approximately 20 ps, recrystallization becomes dominant, and the Ga HCP structure gradually recovers toward stabilization. At 300 K, the wurtzite structure is almost fully recovered after 430 MeV Kr irradiation, whereas only ~95% recovery is achieved following 1171 MeV Ta irradiation, highlighting the more severe structural damage caused by SHIs with higher $S_e$ values. Moreover, a lower Ga HCP fraction at higher temperatures manifests more pronounced damage to the wurtzite GaN structure.

Beyond the recrystallization process, the final ion track morphology reveals permanent structural damage. The evolution of the amorphous/track radius throughout the entire TTM-MD simulation process for GaN irradiated by 430 MeV Kr and 1171 MeV Ta ions at different temperatures are provided in Section 3 of the supplementary materials. Figure 2(a) presents the cross-sectional and longitudinal diagrams of the final ion tracks under 430 MeV Kr irradiation at varying temperatures. Between 300 K and 900 K, almost no ion tracks are observed. When the temperature reaches 1200 K, ion tracks begin to appear with a radius of approximately $0.3 \pm 0.2$ nm. As the temperature continues to rise, the ion tracks become increasingly evident and continuous, accompanied by the formation of discontinuous bubbles. At 1800 K, the continuous ion track radius reaches about $0.8 \pm 0.2$ nm, with the longitudinal dimension of the discontinuous bubbles measuring $2.4 \pm 0.5$ nm. However, under 1171 MeV Ta irradiation, distinct continuous ion tracks emerge even at 300 K, manifesting as discontinuous bubbles with a radius of approximately $1.5 \pm 0.2$ nm and a longitudinal dimension of about $4.1 \pm 0.5$ nm, as depicted in Figure 2(b). As the temperature elevates, the dimensions of the discontinuous bubbles at the continuous ion track center gradually enlarge, particularly with progressive elongation in the longitudinal direction. When the temperature rises to 1500 K, the morphology of the continuous ion tracks transitions from discontinuous bubbles to a continuous channel, with a radius of 2.1 ± 0.2 nm. At 1800 K, the ion tracks formed by continuous channels undergo further expansion, reaching a final radius of approximately 2.3 ± 0.2 nm.

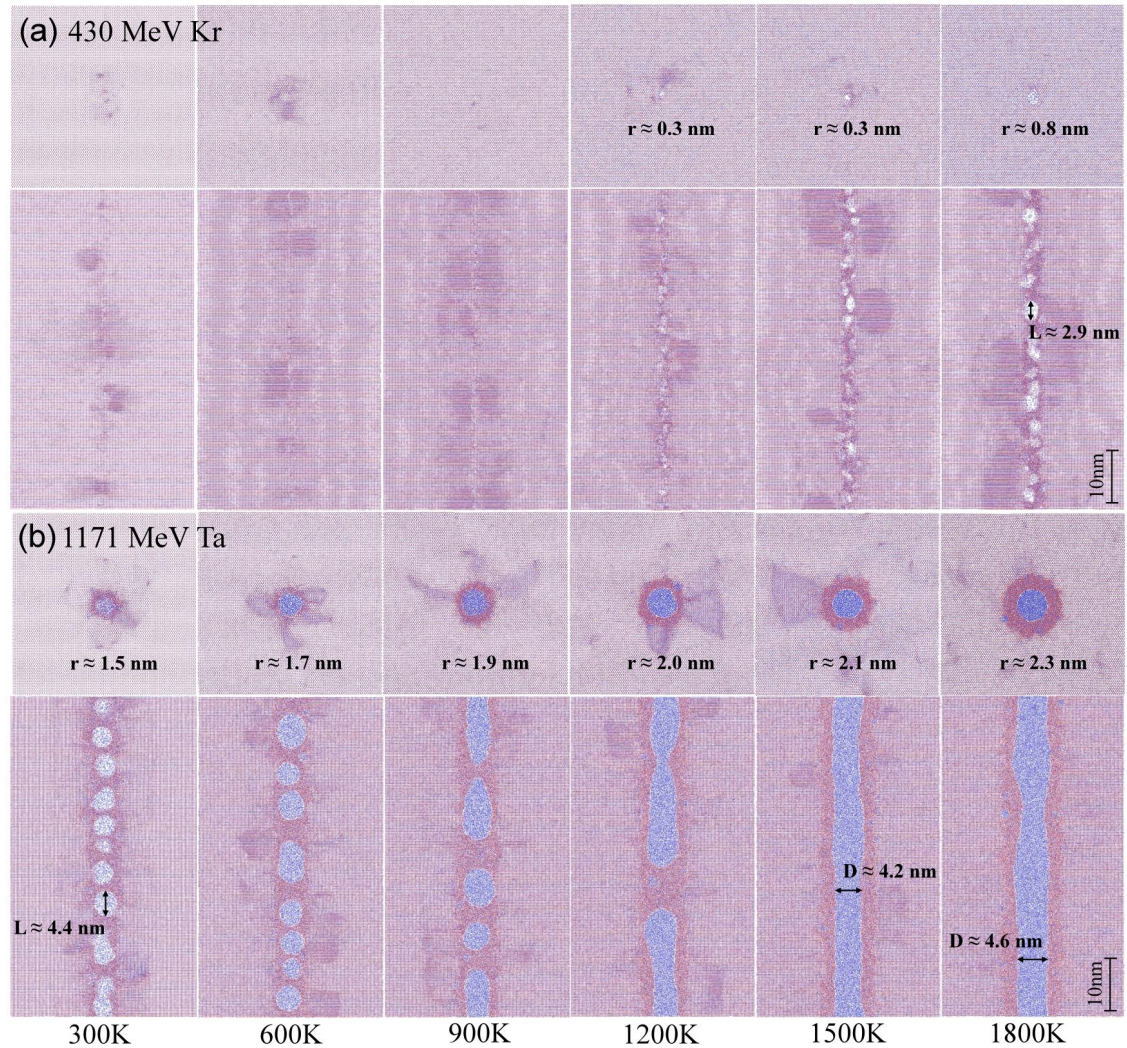

**FIG. 2.** The cross and longitudinal diagrams (25 nm thickness) of final ion tracks induced by (a) 430 MeV Kr and (b) 1171 MeV Ta irradiation.

These phenomena indicate that elevated temperature modulates the morphology of ion tracks in GaN induced by SHI irradiation. Under SHI irradiation with low $S_e$ values, virtually no significant ion tracks are formed at room temperature. Increasing temperature enhances the prominence of ion tracks, leading to an increase in track radius and the generation of discontinuous bubbles. Meanwhile, under high-$S_e$ SHIs irradiation, distinct continuous ion tracks composed of discontinuous bubbles are produced even at room temperature. Subsequent temperature elevation triggers a morphological transition of these continuous ion tracks, evolving from discontinuous bubbles to continuous channels, accompanied by an increase in track radius. Collectively, the ion track morphology exhibits a trend from discontinuous tracks to continuous tracks consisting of discontinuous bubbles, and further to continuous tracks composed of continuous channels. This is consistent with the findings of Zainutdinov *et al.*, who reported an increase in ion track size with rising temperature in SiC[18].

The formation of ion tracks is typically accompanied by intense atomic displacement and rearrangement within the material. To reveal the microscopic mechanism of nanoscale bubble formation along the ion track, the radial distribution function (RDF) of N–N atoms in GaN under 1171 MeV Ta irradiation at 300 K and 1800 K was analyzed and presented in Figure 3(a). At 0 ps, the crystal structure is intact, and the characteristic bond length of N–N atoms corresponds to the lattice constant of the wurtzite GaN, approximately 3.2 Å. After a prolonged melting and recrystallization process, at the final stage of 750 ps, the RDF of N–N atoms corresponding to the wurtzite GaN structure at 300 K largely recovers to its initial value, whereas it significantly decreases at 1800 K. Concurrently, the N–N RDF gradually becomes pronounced at approximately 1.1 Å, a bond length characteristic of $N_2$ molecules[23]. This indicates that SHI irradiation generates $N_2$, with its distribution schematically shown in Figure 3(b), being almost entirely located within bubbles at the ion track center, including both discontinuous bubbles and continuous channels. With respect to Ga atoms, a considerable number of Ga atoms accumulate around the ion track. RDF analysis of Ga–Ga atoms in the atomic aggregation region illustrates that at both 300 K and 1800 K, a double peak corresponding to Ga–Ga bonds appear at approximately 2.8 Å and 3.2 Å, as displayed in Figures 3(c) and 3(d). The 2.8 Å peak corresponds to the characteristic bond length of liquid Ga[24,25], while the 3.2 Å peak corresponds to the Ga–Ga bond length in the wurtzite GaN, i.e., its lattice constant[26].

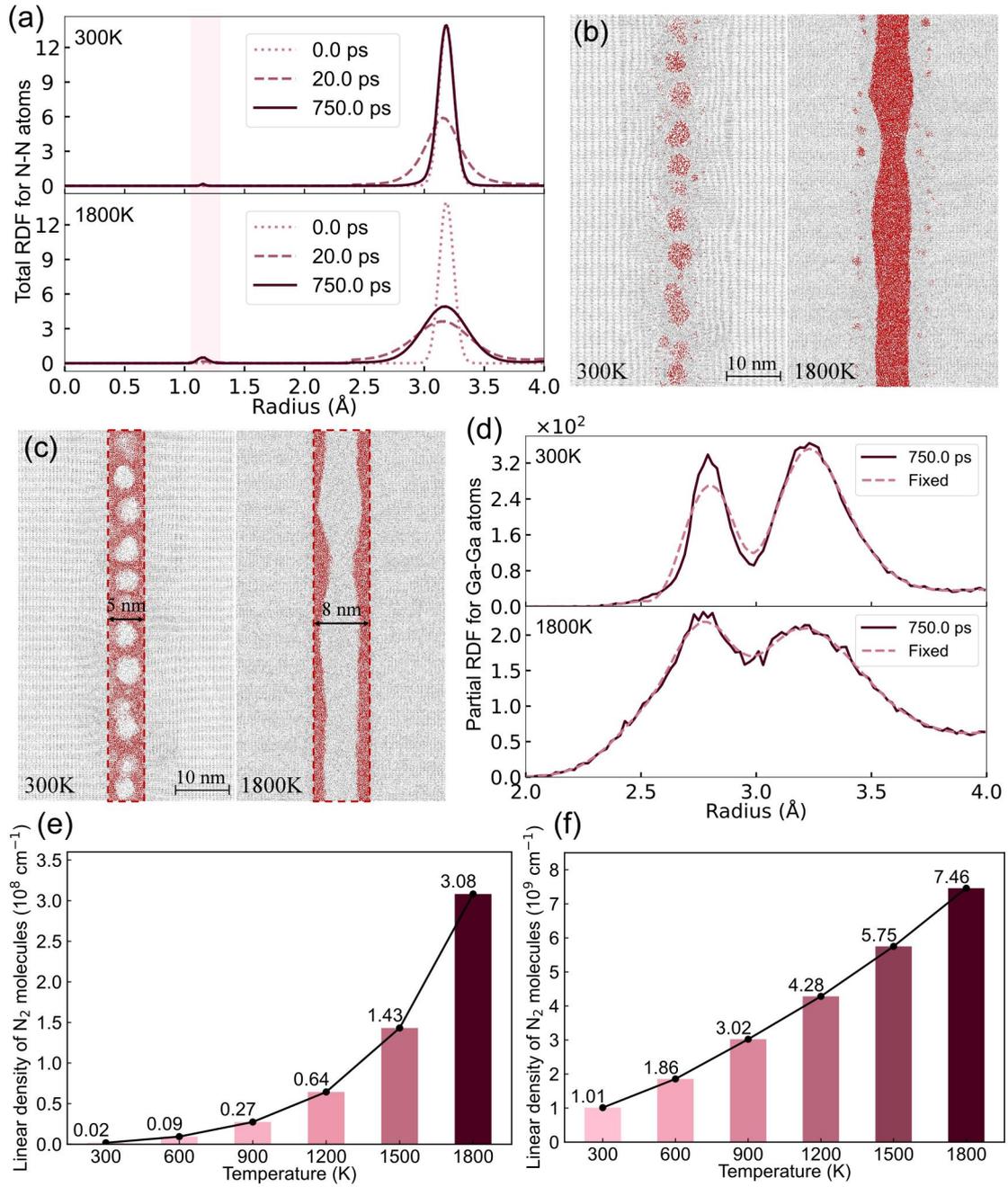

**FIG. 3.** Under 1171 MeV Ta irradiation: (a) Total radial distribution function (RDF) of N-N atoms at 300K and 1800K; (b) Spatial distribution of $N_2$ molecules (red spheres); (c) Spatial distribution of Ga atoms (red spheres) in a selected region near the ion track; (d) Partial RDF of Ga–Ga atoms within the red boxed region in (c). Linear density of $N_2$ molecules under (e) 430 MeV Kr and (f) 1171 MeV Ta irradiation at different temperatures.

This phenomenon suggests that the extreme thermal spike generated in the central region by SHI impact causes instantaneous melting, decomposing the wurtzite GaN structure into liquid Ga and $N_2$ molecules. However, the recrystallization process cannot fully restore the original crystal structure, leaving residual decomposed Ga clusters and $N_2$ molecules at the ion track center. $N_2$ molecules predominantly accumulate within bubbles at the ion track center, including both discontinuous ones and continuous channels. Ga clusters and recrystallized wurtzite GaN primarily aggregate around the periphery of these bubbles, effectively encapsulating $N_2$ molecules within the ion track center, which is generally consistent with experimental observations[10]. Figures 3(e) and 3(f) show the linear density of generated $N_2$ molecules as a function of temperature. Evidently, higher temperature triggers more severe decomposition of the wurtzite GaN structure, resulting in a greater number of $N_2$ molecules at the ion track center. Under SHI incidence with a lower $S_e$ value, the increase in the linear density of generated $N_2$ molecules becomes more pronounced with rising temperature.

Furthermore, the internal extended defects in GaN gradually increase under SHI irradiation, with atomic dislocations being the most prominent. Figure 4(a) presents the total dislocation density in GaN irradiated by 430 MeV Kr and 1171 MeV Ta at different temperatures. The predominant dislocation types are 1/3 <-1 1 0 0> and 1/3 <1 1 -2 0> (see supplementary materials Section 4). At the same temperature, SHI irradiation with higher $S_e$ values produces a significantly greater total dislocation density. However, the dislocation density does not exhibit a monotonic increase with rising temperature, and the temperature corresponding to the maximum dislocation density varies with irradiation conditions. For 430 MeV Kr irradiation, the maximum total dislocation density occurs at 1500 K, whereas for 1171 MeV Ta irradiation, it appears at 1200 K. The reduction in total dislocation density at 1800 K partially confirms that high temperatures promote defect annealing, thereby mitigating the damage retention caused by SHI irradiation[3,27,28].

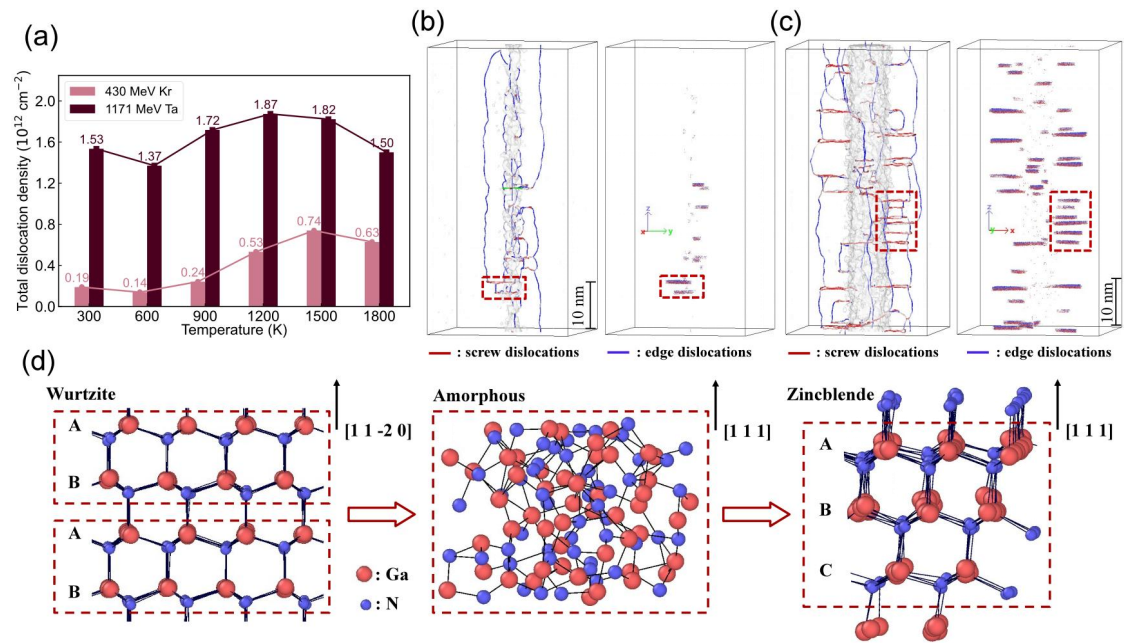

**FIG. 4.** (a) Total dislocation density as a function of temperature. (b)(c) The morphology corresponding to the maximum total dislocation density in (a) and associated irradiation-induced zincblende nanodomains: (b) 430 MeV Kr irradiation at 1500 K; (c) 1171 MeV Ta irradiation at 1200 K. (d) Transformation process of GaN from the wurtzite structure to the zincblende structure.

Existing research has reported that different types of dislocations, such as edge and screw dislocations, exert significantly distinct effects on carrier transport[29]. Screw dislocations can induce electron leakage and current collapse under high voltages, markedly increasing the susceptibility of GaN devices to single-event burnout (SEB), whereas edge dislocations may enhance device performance during high-voltage and high-frequency operation. Figures 4(b) and 4(c) display schematics of dislocations under 430 MeV Kr and 1171 MeV Ta irradiation at the temperatures corresponding to the maximum dislocation densities from Figure 4(a). Remarkably, the dislocations concentrate around the amorphous core of the central ion track. Compared with irradiation by the lower-$S_e$ 430 MeV Kr ions, the dislocations generated under higher-$S_e$ 1171 MeV Ta irradiation are considerably more complex. Notably, the fraction of screw dislocations increases significantly, thereby substantially elevating the risk of GaN device failure[22,30,31].

In addition, a substantial number of GaN structures with perfect face-centered cubic (FCC) configuration are observed around the ion tracks, which has not been documented in previous studies. Structural characterization demonstrates that such FCC structures are a metastable phase of GaN, the zincblende phase. Figures 4(b) and 4(c) show the distribution of irradiation-induced zincblende nanodomains corresponding to the maximum dislocation density under 430 MeV Kr and 1171 MeV Ta irradiation, respectively. Clearly, these zincblende nanodomains spatially coincide with the regions where screw dislocations form. A schematic diagram of the transition process from the wurtzite structure to the zincblende structure within the red box is presented in Figure 4(d). From a crystallographic perspective, the wurtzite structure is based on a HCP framework, with its Ga atomic layers exhibiting an "ABAB" bilayer periodicity along the [11-20] crystallographic direction. In contrast, the zincblende structure is based on a FCC framework, with its Ga atomic layers displaying an "ABCABC" trilayer periodicity along the [111] direction[32,33]. When SHI irradiates the GaN crystal, the enormous deposited energy induces ultrafast melting of the material surrounding the ion trajectory, driven by transient extreme high-temperature and high-pressure thermal spikes. The subsequent cooling process essentially involves the epitaxial recrystallization of liquid GaN. However, this ultrafast melting-quenching process deviates from thermodynamic equilibrium. At extremely high cooling rates, some atoms in the molten region lack sufficient time to relax into the most stable wurtzite structure and instead form the slightly higher energy zincblende structure. Since zincblende GaN possesses distinct bandgap and electronic properties, this is speculated to be one of the key reasons why screw dislocations can induce electron leakage in GaN devices.

4 Conclusion

This study systematically elucidates the temperature-dependent microstructural evolution of ion track morphology and the associated defect generation mechanisms in GaN under SHI irradiation through multi-scale simulations. The results demonstrate that elevated temperatures significantly influence ion track morphology. As temperature increases, the lower $S_e$ value of 430 MeV Kr irradiation promotes the formation of continuous ion tracks and exacerbates bubble generation, whereas the higher $S_e$ value of 1171 MeV Ta irradiation transforms discontinuous bubbles within continuous ion tracks into continuous channels, accompanied by further expansion of the continuous ion track. Meanwhile, SHI irradiation decomposes the wurtzite GaN into liquid Ga and $N_2$ molecules along the ion track, with Ga clusters and recrystallized wurtzite GaN accumulating at the periphery to effectively encapsulate $N_2$ molecules within central bubbles. Atomic-scale analysis further reveals that irradiation-induced zincblende nanodomains form around the ion tracks, with screw dislocations exhibiting strong spatial correlation with these domains, creating additional leakage current pathways, thereby increasing the probability of SEB. These findings clarify the fundamental mechanisms governing defect formation under radiation-induced temperature effects, establishing a theoretical foundation for predicting radiation damage and optimizing radiation-hardened GaN devices for extreme environments.

**SUPPLEMENTARY MATERIAL**

See the supplementary material for detailed descriptions of the two temperature model simulations and visualization results, including the introduction of TTM equation parameters, the presentation of TTM simulation results, the evolution of ion tracks, and schematic illustrations of dislocations.

**Author Contributions**

Conceptualization, Jiayu Liang; methodology, Wenlong Liao; software, Shaowei He; validation, Tan Shi; formal analysis, Hang Zang; investigation, Yonghong Li and Xiaojun Fu; resources, Chuanjian Yao; writing—original draft preparation, Jiayu Liang; writing—review and editing, Chaohui He and Huan He; visualization, Jianan

Wei; supervision, Huan He; funding acquisition, Huan He. All authors have read and agreed to the published version of the manuscript.

**ACKNOWLEDGMENT**

Financial support for this work was provided by the National Natural Science Foundation of China (NSFC) (No. 12405298), China Postdoctoral Science Foundation funded project (No. 2024M762610), and Innovation Scientific Program of CNNC.